\documentclass[%
 reprint,
 amsmath,amssymb,
 aps,
]{revtex4-2}

\usepackage{graphicx}
\usepackage{dcolumn}
\usepackage{bm}
\usepackage{lipsum}
\usepackage{xcolor}

\begin{document}

\preprint{APS/123-QED}

\title{Propagation of Partially Coherent Light in non-Hermitian Lattices}

\author{P. A. Brand\~ao}
\email{paulo.brandao@fis.ufal.br}
\author{J. C. A. Rocha}%
\affiliation{%
 Instituto de F\'isica, Universidade Federal de Alagoas, Macei\'o, 57072-900, Brazil.
}%

\date{\today}

\begin{abstract}
Band theory for partially coherent light is introduced by using the formalism of second-order classical coherence theory under paraxial approximation. It is demonstrated that the cross-spectral density function, describing correlations between pairs of points in the field, can have bands and gaps and form a correlation band structure. The propagation of a partially coherent beam in non-Hermitian periodic structures is considered to elucidate the interplay between the degree of coherence and the gain/loss present in the lattice. We apply the formalism to study partially coherent Bloch oscillations in lattices having parity-time symmetry and demonstrate that the oscillations can be sustained in such media but they are strongly dependent upon the spatial correlations of the beam. A transition between breathing and oscillating modes is shown to be induced by the degree of spatial coherence.
\end{abstract}

\maketitle



\textit{Introduction - } It has been recognized for a long time that the coherence of optical wavefields is one of the main features that dictates how the field will evolve and interact with matter \cite{beran1964theory,mandel1995optical,goodman2015statistical,wolf2007introduction,korotkova2017random,RevModPhys.37.231}. Unfortunately, statistical optics occupy a rather small active portion of researchers' interest, when compared to the fully coherent theories of wave propagation, despite its wide applicability and direct connection to observable quantities \cite{wolf1954optics}. The importance of considering the optical coherence properties is reflected in the numerous developments in technological applications, such as optical coherence tomography \cite{Huang1178}, which enabled non-invasive cross-section imaging of biological tissues, trapping of dielectric particles \cite{Aunon:12}, with the advantage of using beams with low intensity such that biological samples are not damaged, photovoltaics \cite{doi:10.1080/09500340.2017.1363918,lerner2015coherence}, where the coherence properties of sunlight are investigated along with its harvesting at the earth's surface, to cite a few. For an excellent and very readable account on the technological applications of optical coherence theory, the reader can consult Ref. \cite{KOROTKOVA202043}. 
 
 Since the interaction between radiation and matter is ubiquitous in nature, each development of a novel class of dielectric materials is followed by the investigation of the dynamics of light propagation through such media. In this view, there has been enormous interest in so-called non-Hermitian materials, which represent another class of dielectrics that can give to or take energy from the optical wave in a controlled manner \cite{ruter2010observation,guo2009observation}. Apparently, the most successful subclass of non-Hermitian materials that has been proven to give new and unique optical effects is the subclass having the property of Parity-Time (PT) symmetry \cite{longhi2018parity}. They are characterized by the fact that the exchange of energy between radiation and matter is carried out in a balanced way, so that the gain always balances the loss. Curiously, the emergence of this class came from nonrelativistic quantum mechanics, where Bender and Boettcher demonstrated that a quantum Hamiltonian having PT symmetry can generate a real-valued spectrum \cite{bender1998real,bender2018pt,bender2002complex}. 

Despite the large amount of investigation regarding propagation of optical beams in complex materials, the role of the degree of coherence and its relation to PT-symmetric structures has not been fully explored. In fact, the interaction between partially coherent classical light and materials having PT symmetry has just started to be considered in scattering systems \cite{brandao2019non,brandao2019scattering,pinto2020asymmetrical,vieira2020wolf,brandao2021scattering,brandao2021low,korotkova2021light,pires2021scattering,photonics9030140,zhang2022noncentrosymmetric}. These initial results provide strong evidence to the fact that the interplay between gain/loss and the degree of coherence is not trivial and that coherence-induced non-Hermitian effects can be controlled in such systems. Also, the role of the symmetry breaking, which cause an abrupt change in the propagation properties of optical fields, has not been explored in this context. Our objective is to elucidate the obvious advantages of using the formalism of classical coherence to describe beams with partial (spatial) coherence propagating in complex periodic lattices. We hope that the results reported here may place classical coherence theory into a more robust framework in the context of non-Hermitian photonics. As an application, we study a specific system which sustains Bloch oscillations \cite{hartmann2004dynamics,longhi2009bloch} and show how the interplay between non-Hermiticity and the degree of coherence can generate non-trivial dynamics.


\textit{Propagation of partially coherent beams - } Let us begin by considering the stochastic optical field $u(x,z,t)$ in the scalar approximation and assume that it only depends on the $x$ and $z$ spatial coordinates with the $z$ axis denoting the main propagation direction. A monochromatic component of $u(x,z,t)$ with frequency $\omega$ is written as $\psi(x,z,\omega)$ with the dependence on $\omega$ being suppressed from now on. Suppose the beam propagates in a paraxial condition such that $\psi(x,z)$ satisfies the paraxial wave equation (written in scaled units \cite{makris2008beam}),
\begin{equation}\label{paraxial}
    i\frac{\partial \psi(x,z)}{\partial z} = \frac{1}{2}\frac{\partial^2\psi(x,z)}{\partial x^2} + V(x)\psi(x,z),
\end{equation}
where $V(x)$ is the complex potential function related to the refractive index of the structure \cite{longhi2018parity}. In second-order classical coherence, we characterize the beam dynamics through the cross-spectral density function $W(x_1,x_2,z)$ defined as \cite{beran1964theory,mandel1995optical,goodman2015statistical,wolf2007introduction,korotkova2017random,RevModPhys.37.231}
\begin{equation}\label{W1}
    W(x_1,x_2,z) = \Big< \psi^*(x_1,z)\psi(x_2,z) \Big>_{\omega},
\end{equation}
where the average is taken over an ensemble of functions of monochromatic components. Being a field function, the cross-spectral density must evolve according to a differential equation, which is obtained by taking the derivative of \eqref{W1} with respect to $z$ and using \eqref{paraxial},
\begin{equation}\label{wdif}
    i\frac{\partial W}{\partial z} + \frac{1}{2}\left(\frac{\partial^2 W}{\partial x_1^2} - \frac{\partial^2 W}{\partial x_2^2}\right) + \mathcal{V}(x_1,x_2)W = 0,
\end{equation}
where $\mathcal{V}(x_1,x_2) = V^*(x_1) - V(x_2)$ is an effective potential function for the cross-spectral density. A more general differential equation has been derived to describe the evolution of partially coherent light in nonlinear media \cite{shkunov1998radiation,buljan2005partially}. The partial differential equation \eqref{wdif} indicates that if the optical field $\psi(x,z)$ evolves under the potential $V(x)$, its cross-spectral density $W(x_1,x_2,z)$ must evolve under the effective potential $\mathcal{V}(x_1,x_2)$. Apart from the difference between the sign in the second-order derivative in $x$, \eqref{wdif} is essentially a (2+1)-dimensional description of paraxial wave beams $U(x,y,z)$ propagating in the $z$ direction through a material described by the transverse refractive index profile $\mathcal{V}(x,y)$. 

Contrary to the case of monochromatic and fully coherent beams, the cross-spectral density cannot be chosen arbitrarily. It must satisfy the nonnegative definiteness condition $\int W(\mathbf{r}_1,\mathbf{r}_2)f^*(\mathbf{r}_1)f(\mathbf{r}_2)d\mathbf{r}_1d\mathbf{r}_2 \geq 0$ for any choice of well-behaved functions $f(\mathbf{r})$ \cite{mandel1995optical}. A sufficient condition for constructing such genuine correlation functions is obtained by writing \cite{gori2007devising} 
\begin{equation}\label{gori}
    W(x_1,x_2,z) = \int p(v) H^*(x_1,z,v)H(x_2,z,v) dv,
\end{equation}
where $p(v)>0$ and $H(x,z,v)$ is an arbitrary Kernel. Notice that the spectral density $S(x) = W(x,x,z) = \langle |\psi(x,z)|^2 \rangle_{\omega}$ is always positive, independent of the particular form of $H(x,z,v)$. Representation \eqref{gori} also guarantees that the degree of coherence $\mu(x_1,x_2,z)$, defined by
\begin{equation}\label{mu}
    \mu(x_1,x_2,z) = \frac{W(x_1,x_2,z)}{[S(x_1,x_1,z)S(x_2,x_2,z)]^{1/2}},
\end{equation}
satisfies $0 \leq |\mu(x_1,x_2,z)| \leq 1$ such that it can be used to characterize the spatial coherence of the field between positions $(x_1,z)$ and $(x_2,z)$. In dealing with the propagation of paraxial beams, we choose $H(x,z,v) = \int \tilde{w}(k,z,v)e^{ikx}dk$ such that the representation for $W(x_1,x_2,z)$ acquires the interesting form
\begin{equation}\label{w12rep}
    W(x_1,x_2,z) = \iint dk_1 dk_2 w(k_1,k_2,z)e^{i(k_2 x_2 - k_1 x_1)},
\end{equation}
where $w(k_1,k_2,z) = \int \tilde{w}^*(k_1,z,v)\tilde{w}(k_2,z,v)p(v)dv$.  

\textit{Band theory for partially coherent light - } Considering a periodic material described by the complex-valued potential function $V(x) = \alpha\sum_{n=-\infty}^{\infty}c_ne^{2\pi i nx/L}$, where $c_n$ and $\alpha$ are real-valued parameters, $L$ is the lattice period and after substituting \eqref{w12rep} into \eqref{wdif}, we obtain the differential equation governing the evolution of $w(k_1,k_2,z)$:
\begin{equation}\label{spectraldif}
\begin{split}
i\frac{dw(k_1,k_2,z)}{dz}& = \frac{1}{2}(k_1^2 - k_2^2)w(k_1,k_2,z) \\
&+ \alpha\sum_{n=-\infty}^{\infty}c_n\left[ w\left( k_1,k_2 - \frac{2\pi n}{L},z \right) \right. \\
&\left. - w\left( k_1-\frac{2\pi n}{L},k_2,z \right)\right].
\end{split}
\end{equation}
Notice how the definition of $w(k_1,k_2,z)$ implies the symmetry $w(k_1,k_2,z) = w^*(k_2,k_1,z)$ which is already satisfied by \eqref{spectraldif}. This equation indicates that the lattice couples the $k$-space cross-spectral density of the incident beam into discrete regions in the $(k_1,k_2)$ plane during propagation. We call attention to the fact that this is very similar to the standard theory of coherent waves propagating in two-dimensional periodic lattices, but here this coupling happens in the correlation function while the physical system is intrinsic (1+1)-dimensional. Furthermore, the negative sign present in the Laplacian operator (which is reflected in the $k_1^2 - k_2^2$ term) introduces several changes in the band structure of the system and in the propagation properties of the field, as described next.

\begin{figure}[h!]
    \centering
    \includegraphics[width=\linewidth]{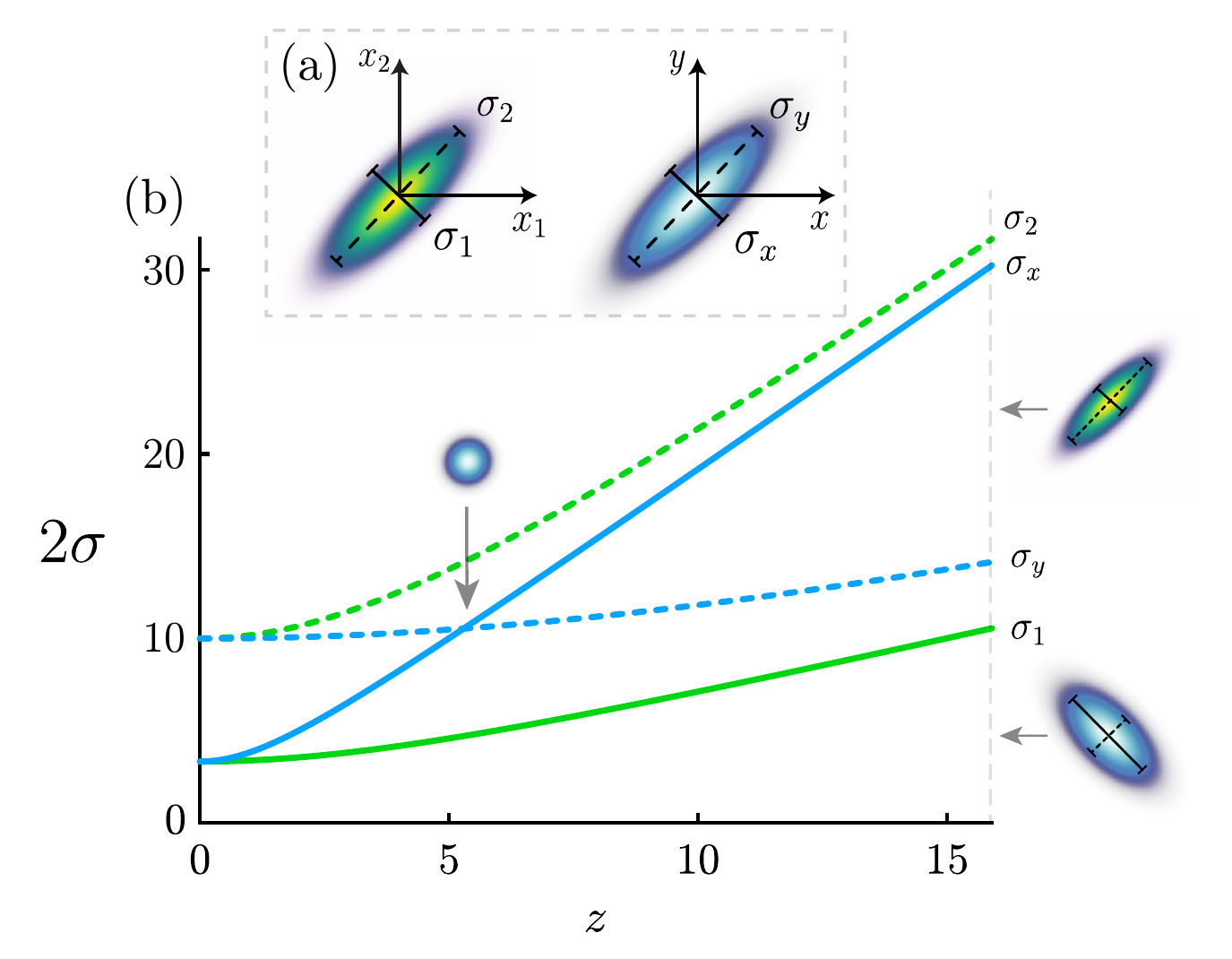}
    \caption{Comparison between the beam waist $\sigma_{1,2}$ of $W(x_1,x_2,z)$ and $\sigma_{x,y}$ of $U(x,y,z)$ in the diagonal positions $x_1\pm x_2=0$ and $x\pm y=0$ during propagation in free-space. (a) Initial profiles at $z = 0$ given by the same function $f(\xi_1,\xi_2,0) = e^{-(\xi_1^2+\xi_2^2)/2\delta^2}e^{-(\xi_1 - \xi_2)^2/2\Delta^2}$ with $W(x_1,x_2,z)$ evolving according to \eqref{wdif} and $U(x,y,z)$ propagating with the same equation but with a plus sign in the Laplacian term (standard paraxial propagation). (b) The position where the blue curves cross indicates that $\sigma_x = \sigma_y$ so that the beam $U(x,y,z)$ becomes symmetrical. This crossing never occurs for $W(x_1,x_2,z)$. Parameters used: $\delta = 10$ and $\Delta = 5$ }
    \label{fig:1}
\end{figure}

In the absence of the lattice, the general solution of \eqref{spectraldif} is given by $w_{\alpha = 0}(k_1,k_2,z) = w_{\alpha=0}(k_1,k_2,0)e^{-\frac{i}{2}(k_1^2 - k_2^2)z}$, where the exponential factor closely resembles the Fourier propagator for (2+1) coherent systems. Figure \ref{fig:1} shows the main difference between the free-space propagation of a monochromatic beam $U(x,y,z)$, evolving under the standard paraxial wave equation, and the cross-spectral density $W(x_1,x_2,z)$, evolving under \eqref{wdif} with $\mathcal{V}=0$, when both are given the same initial asymmetrical field distribution. The figure plots the evolution of the beam widths ($\sigma_{1,2}$ for $W$ and $\sigma_{x,y}$ for $U$) along the two main diagonals of the field ($x_1\pm x_2 = 0$ and $x\pm y=0$), defined as twice the second moment of the intensity distribution \cite{siegman1998maybe}. In the usual situation where the field evolves under the Laplacian operator, the beam suffers a $\pi/2$ rotation along the propagation direction, due to the fact that the smaller waist $\sigma_x$ grows faster than the larger one, $\sigma_y$. This effect does not occur for the evolution given by \eqref{wdif}. In the particular case where the initial beam width is symmetrical, no changes are observed during propagation (apart from the usual diffraction which tends to increase the overall width) in both cases. However, it is important to consider an asymmetrical case because in the partially coherent scenario this is usually the situation of interest, as will be discussed later. 

Consider now the effects of the lattice. Without loss of generality, assume that the potential $V(x)$ has only three nonzero Fourier components, $c_0$, $c_{\pm1}$, and write $c_0 = 1$ and $c_{\pm1} = \frac{1}{2}(1\pm\gamma)$ where $\gamma \geq 0$. The Hermitian configuration with real $V$ is recovered if $\gamma = 0$ and for $\gamma \neq 0$ the potential is PT-symmetric with $\gamma = 1$ being the symmetry breaking point. Let us now try to define a coherence band structure, analogous to the usual band structure of coherent waves, by writing $w(k_1,k_2,z) = p(k_1,k_2)e^{i\beta z}$, where $\beta$ is the coherence eigenvalue, and substituting this form into \eqref{spectraldif}. The coherence band structure for $\alpha = 0$ is seen to be $\beta = \frac{1}{2}(k_2^2 - k_1^2)$, described by a hyperbolic paraboloid, which has a very different topology when compared to the band structure of a two-dimensional coherent beam (described by a paraboloid) due to the minus sign in the expression. 

\begin{figure}[!ht]
    \centering
    \includegraphics[width=\linewidth]{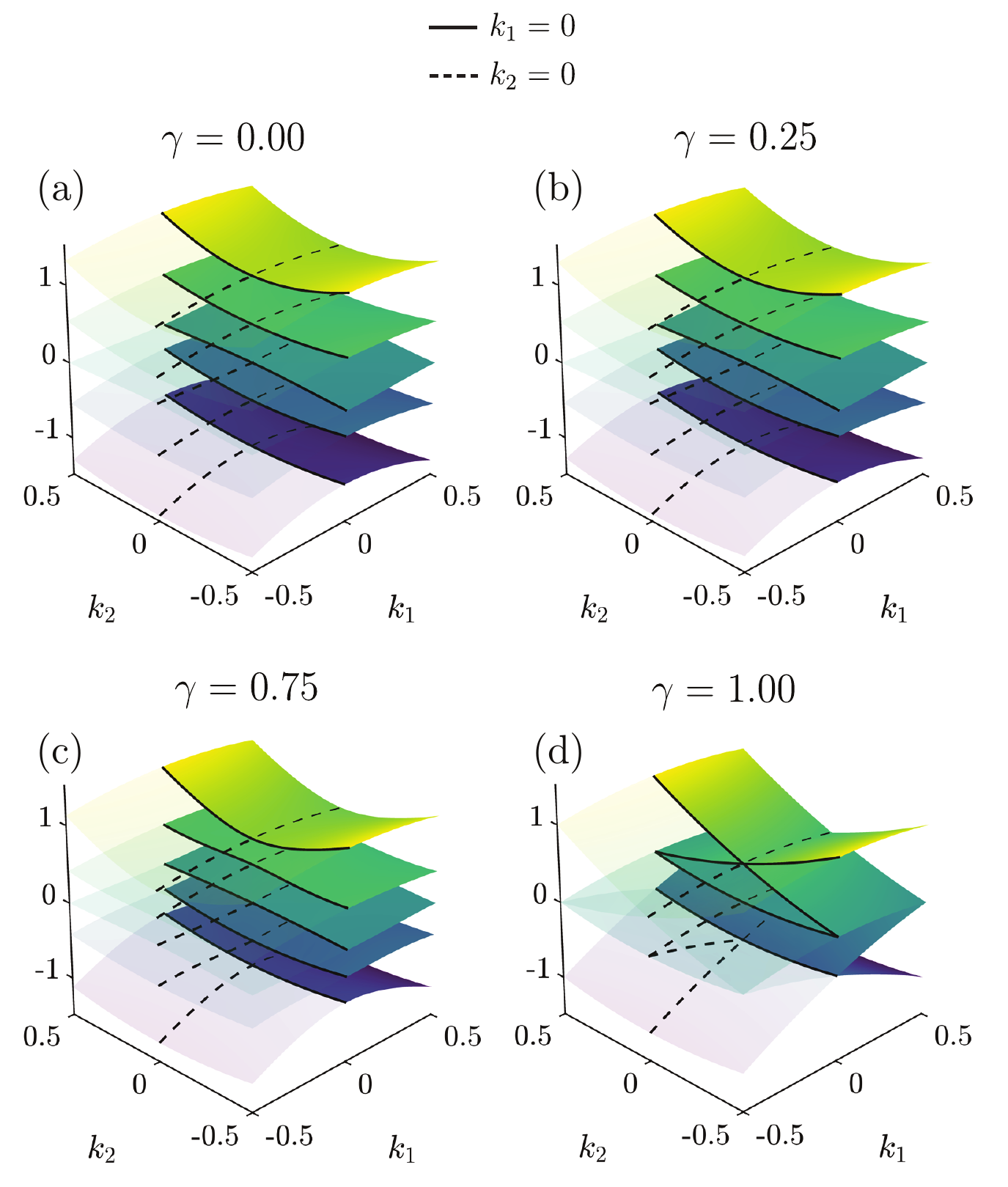}
    \caption{Coherence band structures for partially coherent light in non-Hermitian lattices with (a) $\gamma = 0$ (Hermitian lattice), (b) $\gamma = 0.25$ and (c) $\gamma = 0.75$ (PT-symmetric lattices below threshold), and (d) $\gamma = 1$ (PT-symmetric lattice at the symmetry breaking point). The bands in the last case are identical with the bands in the absence of the lattice $\alpha = 0$. The lattice period is $L = 2\pi$.}
    \label{fig:2}
\end{figure}

This unusual topology has its origin in the fact that, from the symmetry requirement $w(k_1,k_2,z)=w^*(k_2,k_1,z)$, the necessary condition $\beta = 0$  must be satisfied whenever $k_1 = k_2$. This is because $w(k,k,z)$ has to be a real-valued function. No such requirements are needed when considering the usual coherent propagation because $U(x,y,z)$ does not represent a correlation function. Therefore, the necessity of the cross-spectral density to be a genuine correlation profile changes the topology of the coherence band structure. Figure \ref{fig:2} shows the plots of the band structure for several values of $\gamma$. Part (d) of this figure indicates the symmetry breaking point where the band-merging occurs. Clearly, non-Hermiticity can influence the propagation of partially coherent beams in a non-trivial way.

\begin{figure}[!ht]
    \centering
    \includegraphics[width=\linewidth]{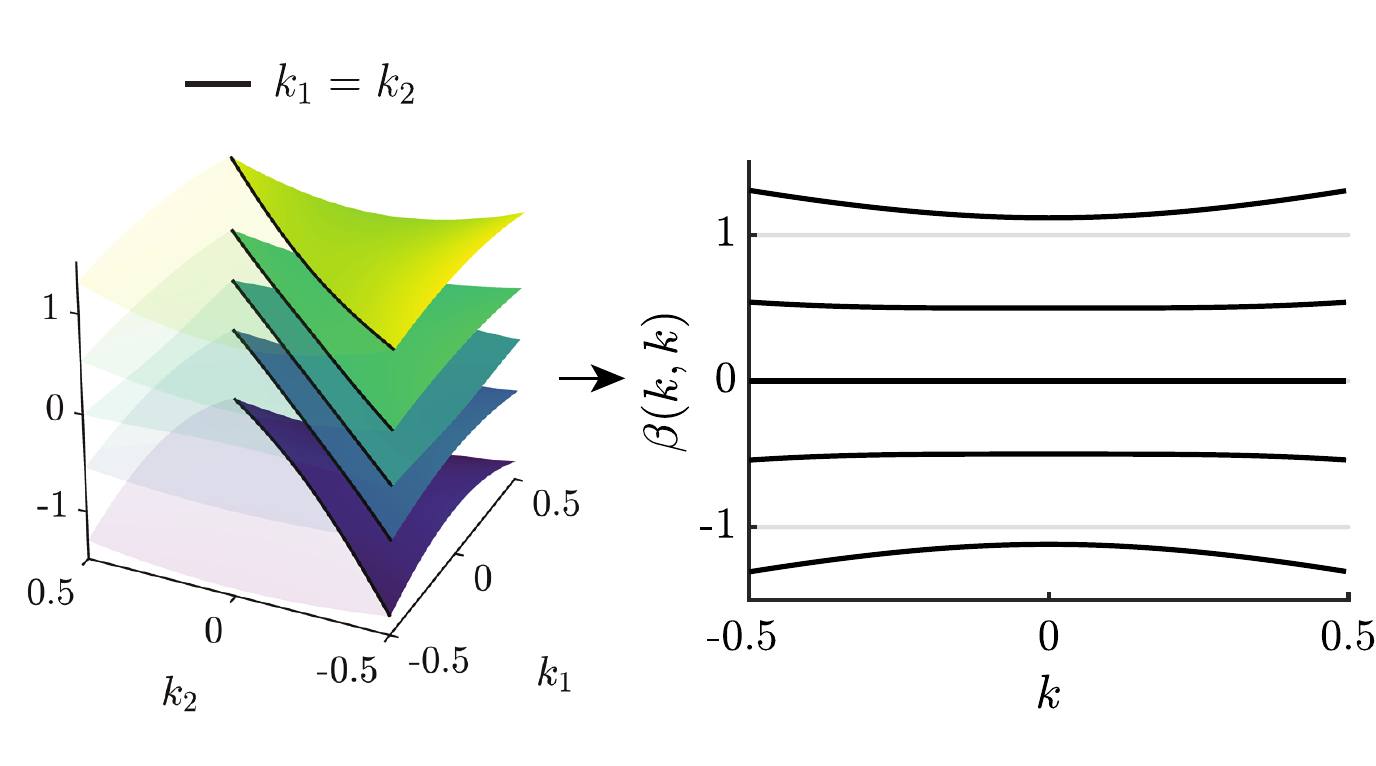}
    \caption{Cross-section along the diagonal ($k_1 = k_2$) of the band diagram for $\gamma = 0.5$. This figure indicates the symmetrical distribution of the positive and negative coherence eigenvalues $\beta$.}
    \label{fig:3}
\end{figure}

In this view, one can still form correlation wavepackets by linear combinations of $p_n(k_1,k_2)e^{i\beta_n z}$ as long as $w(k_1,k_2,z)$ be a real-valued function for $k_1 = k_2$. It is thus necessary that for every positive eigenvalue $\beta_n(k,k)$ present in the wavepacket there is a corresponding negative one $-\beta_n(k,k)$. The band diagram must reflect this symmetry and Figure \ref{fig:3} highlights the eigenvalues along the diagonal $k_1 = k_2$ to confirm these claims. This explains why the bands are symmetrical with respect to $\beta = 0$ along the $k_1 = k_2$ axis. 

\begin{figure}[]
    \centering
    \includegraphics[width=0.9\linewidth]{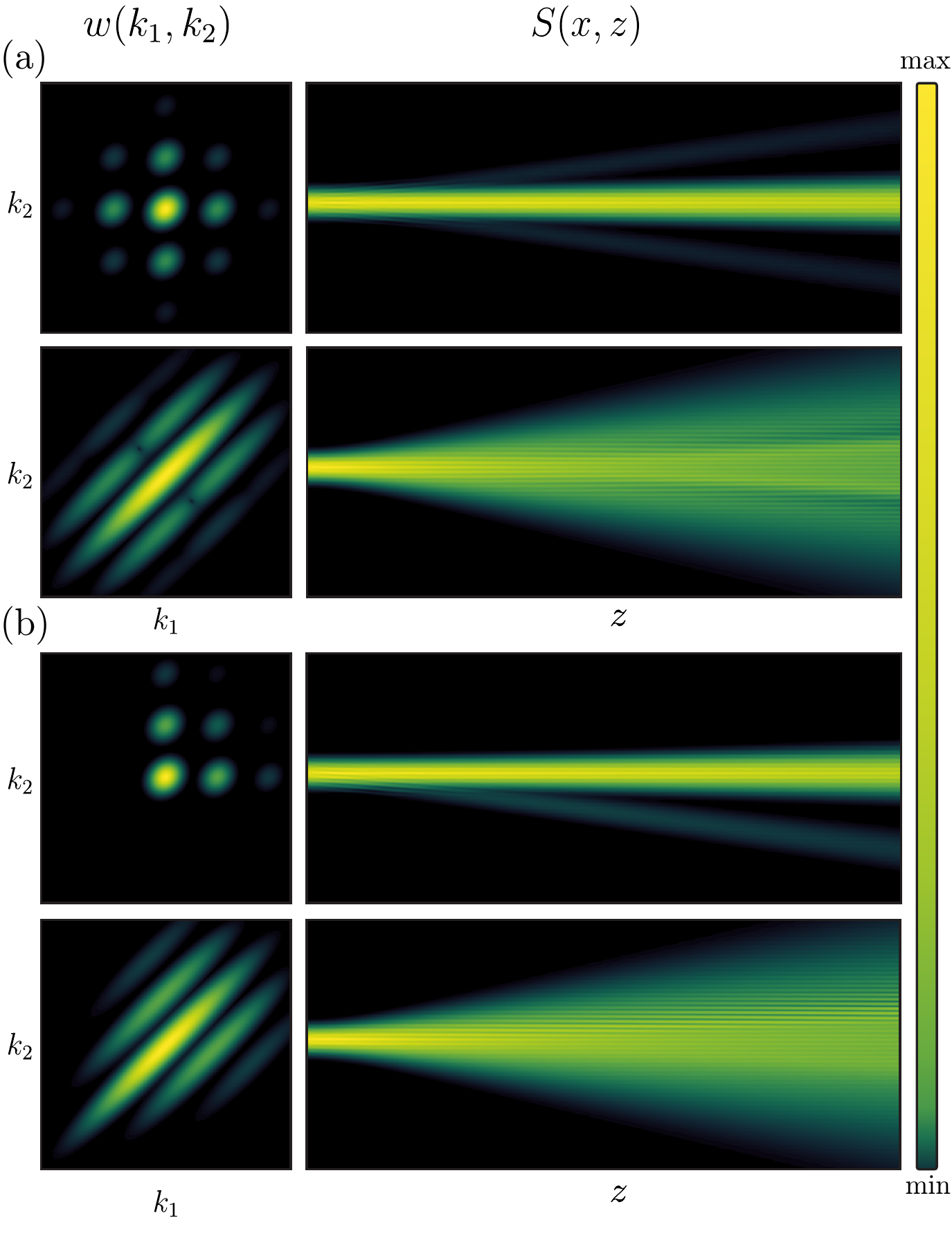}
    \caption{Double refraction of a Gaussian-Schell beam propagating in Hermitian and non-Hermitian lattices. (a) Hermitian lattice with $\gamma = 0$ and (top) $\Delta = 20$, (bottom) $\Delta = 2$. The right panel displays the spectral density. The spectral correlation function $w(k_1,k_2,z)$ couples to the lattice according to \eqref{spectraldif}. (b) Evolution in a non-Hermitian lattice with $\gamma = 1$. In both cases the $k$-space correlation function broadens as the degree of spatial coherence decreases, washing out any non-Hermitian effects on its propagation. The incident beam width remained fixed at $\delta=10$.} 
    \label{fig:4}
\end{figure}

 The significance of the coherence eigenstates $w_n(k_1,k_2,z)$ becomes clear if we consider the $k$-space degree of coherence $\mu_w^{(n)}(k_1,k_2,z) = w_n(k_1,k_2,z)/[w_n(k_1,k_1,z)w_n(k_2,k_2,z)]^{1/2}$, defined in analogy with the spectral degree of coherence, and note that $|\mu_w^{(n)}|$ is independent of $z$, which is the partially coherent version of the stationary states present in the usual band theory for coherent states.

\textit{Beam dynamics - } It is now time to discuss some beam dynamics. Consider a beam having finite transverse extent $\delta$ propagating in the complex lattice $V(x)$. The beam is assumed to be initiated at $z = 0$ and described by the Gaussian-Schell model with the  cross-spectral density $W(x_1,x_2,0)$ given by
\begin{equation}\label{gschell}
    W(x_1,x_2,0) =S_0e^{ -\frac{x_1^2+x_2^2}{2\delta^2} }e^{ -\frac{(x_1 - x_2)^2}{2\Delta^2}},
\end{equation}
 where $\delta$ is related to the beam width at $z = 0$ and $\Delta$ is spatial coherence parameter which controls the statistical correlations between positions $x_1$ and $x_2$. The spectral density is given by a Gaussian function $S(x,0) = W(x,x,0) = S_0e^{ -\frac{x^2}{\delta^2} }$, where $S_0$ is the amplitude of the incident beam. From \eqref{mu} we obtain the spatial degree of coherence, $\mu(x_1-x_2) = e^{ -\frac{(x_1 - x_2)^2}{2\Delta^2}}$. The beam is said to be fully coherent in the limit $\Delta \rightarrow \infty$ ($\mu \rightarrow 1$) where $W(x_1,x_2)$ becomes a separable function, as can be verified. We are mainly interested in the relationship between the coherence parameter $\Delta$ and the non-Hermitian lattice and how they influence the beam propagation through the material.

 For the Gaussian-Schell model \eqref{gschell}, the function $w(k_1,k_2,0)$ can be calculated directly from \eqref{w12rep}. It is given by
 \begin{equation}
     \begin{split}
         w(k_1,k_2,0) &= \frac{2\pi S_0 \delta^2\Delta}{\sqrt{2\delta^2 + \Delta^2}}\exp\left\{\ - \frac{\delta^2}{2(2\delta^2 + \Delta^2)} \right. \\
         &\left.\times \left[(k_1 - k_2)^2\delta^2 + (k_1^2 + k_2^2)\Delta^2  \right] \right\}\ ,
     \end{split}
 \end{equation}
which can be used as the initial condition in \eqref{spectraldif}. However, it is more efficient to solve \eqref{wdif} directly by using split-step Fourier numerical methods, with an absorbing material on the boundaries to prevent reflection of the outgoing modes, and use \eqref{gschell} as the initial condition \cite{pedrola2015beam}. We also compared all numerical results with second-order perturbation theory and obtained an excellent agreement. Figure \ref{fig:4} shows the evolution of the spectral density $S(x,z)$ (right panels), along with $w(k_1,k_2,z)$ (left panels), for a passive $\gamma = 0$ and PT-symmetric $\gamma = 1$ lattice. In the Hermitian case with (large) spatial coherence $\Delta = 20$, we see in the top panel of Fig. \ref{fig:4}(a) the usual Bragg diffraction with symmetrical first-order modes, which reflects the symmetry of the coherence lattice $w(k_1,k_2,z)$. However, as the degree of spatial coherence decreases, the then distinct discrete Bragg modes merge in a symmetric way along the diagonal $k_1 = k_2$ and the overall effect of the Bragg modes is washed out during propagation.

\begin{figure}[]
    \centering
    \includegraphics[width=0.9\linewidth]{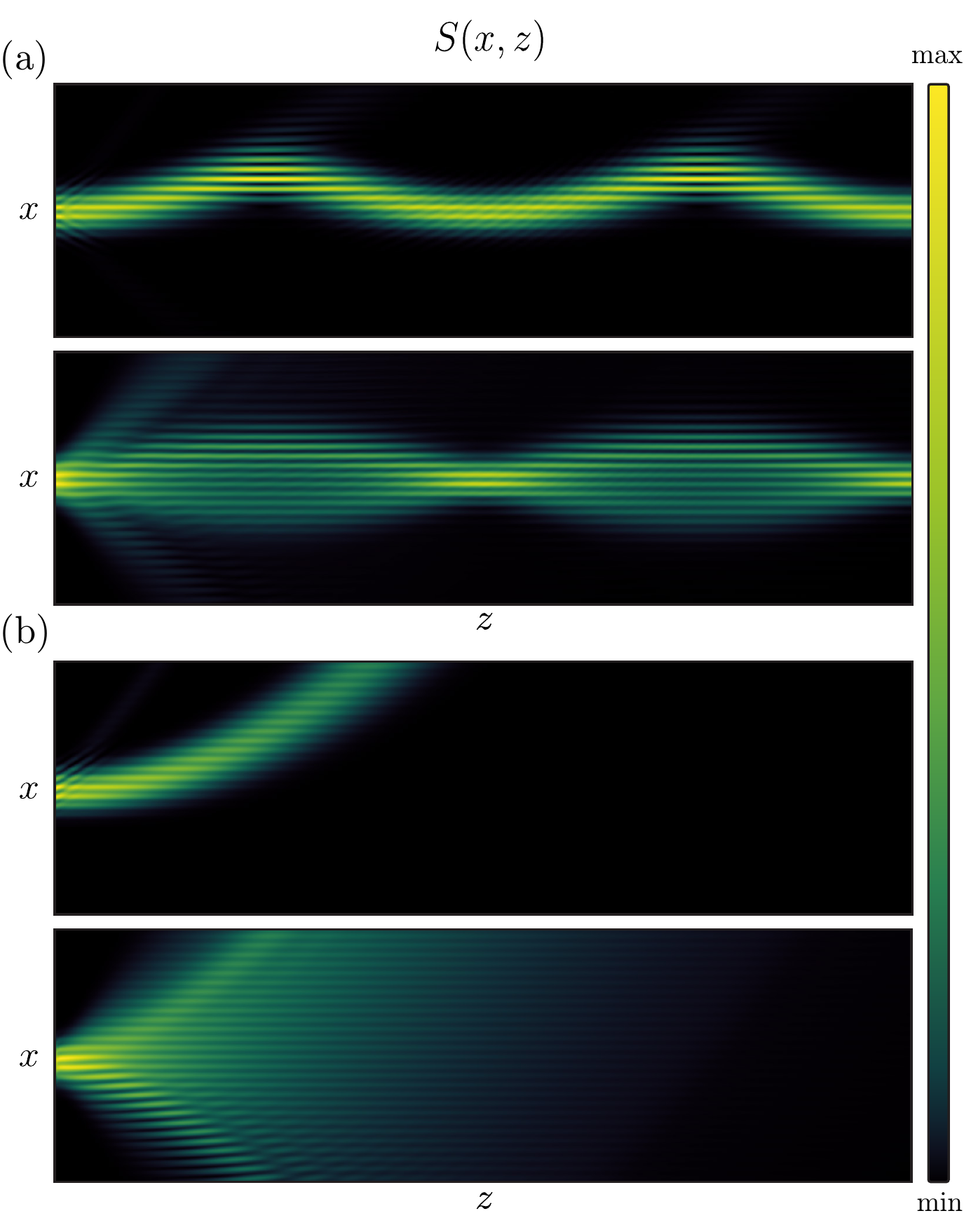}
    \caption{Bloch oscillations with partially coherent light in complex media. (a) Coherence-induced transition between oscillating and breathing modes with $\gamma = 0$, (top) $\Delta = 100$ and (bottom) $\Delta = 2$. (b) Evolution through the lattice at the symmetry breaking point $\gamma = 1$ with (top) $\Delta = 100$, (bottom) and $\Delta = 2$. In both cases the initial beam width is $\delta = 10$, $\beta = -0.2\times10^{-3}$, $L = 2\pi$ and $\alpha = 5\times10^{-3}$.} 
    \label{fig:5}
\end{figure}

In the case of a lattice having gain and loss at the symmetry breaking point $\gamma = 1$, and large coherence, a few of the Bragg modes are not excited and the beam evolution displays the phenomenon of double refraction \cite{makris2008beam}, as can be seen in Fig. \ref{fig:4}(b). However, as $\Delta$ decreases, the spatial correlation induces the creation of new modes in such a way that the observed double refraction is destroyed, albeit in a more asymmetrical way. The curious pattern of diagonal stripes can be understood by noticing that in the $\Delta \rightarrow 0$ limit, the cross-spectral density function must behave as $W(x_1,x_2,z) \propto \delta(x_1 - x_2)$ and from \eqref{wdif} we concluded that $w(k_1,k_2,z)$ depends only on the difference between $k_1$ and $k_2$. Since in this limit we return to the fully coherent case, the line $k_1 = - k_2$ represents the usual excited Bragg modes for the coherent system. We emphasize that the initial beam width is the same in all simulations, so it is indeed the spatial correlation that is exciting new modes.

As the previous results suggest, the broadening in the correlation space as $\Delta$ decreases, i.e., as the beam becomes more spatially incoherent, can lead to a nontrivial dynamics and excitation of a large number of modes for a fixed initial beam width. In order to show a more dramatic example, consider adding a linear ramp to the periodic lattice, thus breaking the periodic symmetry of the material. More quantitatively, let us consider the propagation of the Gaussian-Schell beam \eqref{gschell} through the potential $V(x) = \alpha\sum_{n = -\infty}^{\infty}c_ne^{2\pi i n x /L} + \beta x$, where $\beta$ is a constant parameter, again with $c_0 = 1$, $c_{\pm1} = \frac{1}{2}(1\pm \gamma)$ and $c_n = 0$ ($|n|>1$).

Figure \ref{fig:5} plots the beam evolution by numerically solving \eqref{wdif}. Part (a) displays the Hermitian case with (top panel) large $\Delta = 100$ and (bottom panel) small $\Delta = 2$ spatial coherence for a fixed beam width $\delta = 10$. We observe a coherence-induced transition between oscillating and breathing modes of Bloch oscillations as $\Delta \rightarrow 0$. To explain this, recall that the breathing mode is observed in fully coherent systems as the incident beam width decreases \cite{hartmann2004dynamics}. The initial beam width of our simulation remained fixed at $\delta = 10$. However, in the correlation space $(k_1,k_2)$, the overall area of the beam width (which is composed of $\delta$ and $\Delta$) actually decreases as $\Delta \rightarrow 0$. In other words, less modes are initially excited in the $\Delta \rightarrow 0$ limit. Since it is the spatial correlation that determines the beam evolution, the role of the partial coherence is reflected in the real space propagation of the spectral density $S(x,z) = W(x,x,z)$. 

In the case of a lattice at the symmetry breaking point $\gamma = 1$, \eqref{wdif} is able to describe the usual behavior of an accelerating beam, shown in the top panel of Fig. \ref{fig:5}(b), as if propagating in free-space. In the low coherence regime, shown in the bottom panel, the partially coherent beam spreads strongly, albeit in a more skewed distribution. This suggests that it is possible to control, and even cancel, the accelerating effect present in fully coherent systems.

\textit{Final remarks - } The inclusion of partial coherence in optical systems can generate new dynamics not present in the fully coherent counterparts. In the same way that a real, Hermitian, theory can be extended to the complex, non-Hermitian, domain, a fully coherent theory can be made partially coherent by introducing the cross-spectral density function, as done here. In this respect, every theory published involving propagation of beams through paraxial conditions can be generalized to include the role of the degree of coherence. Our results provide the first step towards an understanding between the role of spatial coherence and complex media in paraxial conditions. 

The authors acknowledge the financial support of Conselho Nacional de Desenvolvimento Cient\'ifico e Tecnol\'ogico (CNPq).

\nocite{*}

\bibliography{apssamp}

\end{document}